\documentstyle[twoside,fleqn,hyperon99_paper]{article}

\newcommand{\qslash}{\mbox{$\not \!q$}}

\newcommand{\AmS}{{\protect\the\textfont2
  A\kern-.1667em\lower.5ex\hbox{M}\kern-.125emS}}

\title{Weak radiative hyperon decays: questioning the basics}

\author{P. \.Zenczykowski \address{Institute of Nuclear Physics,\\
        Radzikowskiego 152, 31-342 Krak\'ow, Poland}%
        }
       
\begin{document}

\begin{abstract}
Main theoretical approaches to weak radiative hyperon decays are 
briefly reviewed. 
It is emphasized that only approaches with great predictive power
should be seriously considered  
when seeking a resolution of the puzzle
presented by observed large negative asymmetry
$\alpha (\Sigma ^+ \rightarrow p \gamma )$. 
In such cases, asymmetry in the
 $\Xi ^0 \rightarrow \Lambda \gamma $ decay is always large while its sign
is positive (negative) if Hara's theorem is violated (satisfied). 
Measuring this asymmetry is therefore crucial 
for determining whether the large value of 
$\alpha ( \Sigma ^+ \rightarrow p \gamma )$ is due to large SU(3) breaking 
or to some deeper reason.
Some arguments suggesting that violation of Hara's theorem might be a feature
of Nature, and hints as to its possible origin are also given.

\end{abstract}

\maketitle

\section{INTRODUCTION}

Weak radiative hyperon decays (WRHD's) present a yet unsolved
problem in low-energy physics of hadronic weak interactions.
The issue first appeared in 1969 when
measurements of the $\Sigma ^+ \rightarrow p \gamma $ decay
asymmetry \cite{Gershwin} gave 
$ \alpha ( \Sigma ^+ \rightarrow p \gamma) = -1.0 ^{+0.5}_{-0.4}$.
This value was not in agreement with expectations based on Hara's theorem
\cite{Hara}, according to which the asymmetry in question should be small.
Problems with Hara's theorem have plagued the issue of WRHD's ever since.

Hara's theorem states that the parity-violating amplitude $A$ for 
the  $\Sigma ^+ \rightarrow p \gamma $ 
(and $\Xi ^- \rightarrow \Sigma ^- \gamma $) decay
should vanish in exact flavor SU(3).
In reality, SU(3) is broken of course. 
However, if the parity-conserving amplitude $B$ is not small ($A \ll B$), 
one expects the asymmetry
 $\alpha = 2 A B/(A^2+B^2) \approx 2A/B$ to be small (i.e. not larger than
 ca $\pm 0.2$). 
The theorem follows if hadrons are described by an SU(3)-symmetric
gauge- and $CP$- invariant local field theory.
Although these assumptions (with the exception of SU(3), of course) 
are fundamental,
one should note here that the very year the theorem was proved (1964),
significant changes 
in our knowledge about these assumptions occurred.
Thus, 1) it was proposed that SU(3) should follow from the underlying 
quark model,
2) violation of CP invariance was experimentally observed, and 3) the first
paper pinning down the nonlocal nature of quantum physics appeared. 
These changes should be kept in mind when considering possible
theoretical reasons for the experimentally found departure 
from expectations based on Hara's theorem.

At present, we know that asymmetry in the $\Sigma ^+ \rightarrow p \gamma $
decay is large. The PDG average \cite{PDG} is
$\alpha (\Sigma ^+ \rightarrow p \gamma )= -0.76 \pm 0.08$, with two main
experimental results contributing equal to
$-0.86 \pm 0.13 \pm 0.04 $ \cite{KEK},
and $-0.72 \pm 0.086 \pm 0.045 $ \cite{Foucher} respectively.
Therefore, the situation is quite disturbing since with
one baryon in the initial state and one baryon in the final state 
(and thus lacking strong interactions in the final state),
the WRHD's are fairly clean transitions, 
similar to the semileptonic ones or to magnetic moments.
With the only WRHD-specific complication being joint appearance 
of weak and electromagnetic interactions,
a fairly precise theoretical description of WRHD's should be 
then possible.

\section{SIZE OF DATA BASIS AND RELIABILITY OF CONCLUSIONS}

Although we may sum up the experimental findings by saying 
that expectations based on Hara's
theorem are strongly violated, 
we cannot draw any deeper conclusions as to the origin of the effect.
In this respect the situation is similar to what might have happened if we had
measured the magnetic moment of proton to be $\mu _p \approx 2.79$ but had
not known anything about magnetic moments of other ground-state baryons.
Although we might have stated then that 
large correction to the Dirac value of proton magnetic moment is present, 
no conclusions concerning the {\em symmetric} 
nature of flavor-spin wave functions 
(and hence color) would have been possible.  
Drawing such conclusions requires (at the very least) measuring 
the ratio $\mu _n/\mu _p$ 
which is $-2/3$ ($-2$) for symmetric (antisymmetric)
spin-flavor wave functions.
The lesson is that large asymmetry observed in the
$\Sigma ^+ \rightarrow p \gamma$ decay must be analysed together with
data and theory on the remaining WRHD's, i.e.
$\Lambda \rightarrow n \gamma $,
$\Xi ^0 \rightarrow \Lambda \gamma $,
$\Xi ^0 \rightarrow \Sigma ^0 \gamma $, as well as
$\Xi ^- \rightarrow \Sigma ^- \gamma $ and
$\Omega ^- \rightarrow \Xi ^- \gamma $.
Of these, the first three turn out to be particularly important.
Present experimental data \cite{PDG} are gathered in Table 1.

\begin{table}
\caption{Data  }
\begin{tabular}{lrr}
\hline
Decay & Asymmetry &Br. ratio$\cdot 10^3$\\
\hline
$\Sigma ^+ \rightarrow p \gamma$&$-0.76\pm0.08$&$1.23\pm0.06$\\

$\Lambda \rightarrow n \gamma $&&$1.75\pm0.15$\\
$\Xi ^0 \rightarrow \Lambda \gamma $&$+0.43\pm0.44$&$1.06\pm0.16$\\
$\Xi ^0 \rightarrow \Sigma ^0 \gamma $&$-0.65\pm0.13^{(*)}$&$3.6\pm 0.4$\\
$\Xi ^-\rightarrow \Sigma ^- \gamma$ &$+1.0\pm 1.3 $&$0.127\pm 0.023 $\\
$\Omega ^- \rightarrow \Xi ^- \gamma $&&$<0.46$\\
\hline
$^{(*)}$ Ref. \cite{this}
\end{tabular}
\end{table}

Following the successes of the description of
semileptonic decays and magnetic
moments with the help of one (or two) parameters in each case, 
one may reasonably expect that the puzzle of an apparent violation of 
Hara's theorem in $\Sigma ^+ \rightarrow p \gamma $ 
will be  resolved successfully if 
all radiative decays are well described with the help of an approach
using a very small number of parameters.
In other words, we need an approach which accurately predicts experimental
branching ratios and asymmetries, with errors below 20\%.
Description of asymmetries will provide here a particularly incisive test. 
When such an approach akin to the quark model description of baryon
magnetic moments is available,  its further and deeper analysis  
should be attempted.

\section{THEORY - GENERAL}
\subsection{Hara's theorem}
By using local field theory at hadron level,
Hara's theorem may be obtained as follows.
The most general parity-violating electromagnetic current
may be written as:
\begin{equation}
\label{generalcurrent}
j^{\mu}_{5,kl}=j^{(1)\mu}_{5,kl}+j^{(2)\mu}_{5,kl}
\end{equation}
where $k,l$ are baryon indices,
\begin{equation}
\label{j1}
j^{(1)\mu}_{5,kl}=g_{1,kl}(q^2) \overline{\psi}_k
(\gamma ^{\mu} - q^{\mu} \qslash /q^2)\gamma _5 \psi _l,
\end{equation}
and
\begin{equation}
\label{j2}
j^{(2)\mu}_{5,kl}=g_{2,kl}(q^2) \overline{\psi}_k
(i \sigma ^{\mu \nu} \gamma _5 q_{\nu}) \psi _l.
\end{equation}
Hermiticity and CP invariance of $A\cdot j_5$ require
\begin{equation}
\label{g1}
g_{1,kl}=g_{1,lk} 
\end{equation}
and
\begin{equation}
\label{g2}
g_{2,kl}=-g_{2,lk}
\end{equation}
with $g_{i,kl}$ real.

Hara's theorem is obtained when hadron indices $k,l$ are replaced with
$\Sigma ^+$,$p$.
Since no {\em exactly} massless hadron exists, there cannot 
be a pole at $q^2=0$.
Consequently, $g_{1,kl}(q^2)$ must be proportional to $q^2 $.
Therefore, real transverse photons, for which $q^2=q\cdot A =0$,
interact with the $j^{(2)}$ current only.
Now, under $s \leftrightarrow d $ interchange, $\Sigma ^+ = uus$
goes into $p = uud$ and vice versa.  Thus, in exact SU(3) we must have
$g_{2,\Sigma ^+p} = g_{2,p\Sigma ^+}$.
Since $g_{2,\Sigma^+p}$ is simultaneously 
symmetric and antisymmetric (c.f. Eq.(\ref{g2})),
it must vanish.
(We might have e.g. $g_{2,kl}\propto (m_k-m_l)$).
If, for some reason, $g_{1,\Sigma ^+ p}$ were not equal to $0$,
Hara's theorem might be violated.

\subsection{Quarks}
Any acceptable approach to WRHD's must take into account the fact that
baryons are composites made of quarks. 
From the point of view of essentially any quark-inspired model, 
the WRHD's may be divided into two groups.
The first group consists 
of decays  
arising solely from such transitions in which a single quark 
undergoes a weak transition and radiates a photon.
This occurs e.g. for  $\Xi ^- \rightarrow \Sigma ^- \gamma $ and 
$\Omega ^- \rightarrow \Xi ^- \gamma $.
The other group involves more complicated two-quark processes 
$su \rightarrow ud \gamma $ as well.
This group contains decays
$\Sigma ^+ \rightarrow p \gamma $,
$\Lambda \rightarrow n \gamma $,
$\Xi ^0 \rightarrow \Lambda \gamma $, and
$\Xi ^0 \rightarrow \Sigma ^0 \gamma $.

Assuming that WRHD's are dominated by single-quark transitions, one
can estimate the branching ratio of decay 
$\Sigma ^+ \rightarrow p \gamma $ using 
that of $\Xi ^- \rightarrow \Sigma ^- \gamma $ \cite{GilmanWise}.
Since the latter is experimentally very small (cf. Table 1),
one calculates that single-quark transition may contribute only around
1\% to the experimentally observed $\Sigma ^+ \rightarrow p \gamma $ 
branching ratio.
Thus, it is the two-quark transition $su \rightarrow ud \gamma $
which dominates the  $\Sigma ^+ \rightarrow p \gamma $ decay.
Its properties should be accessible from detailed studies
of the remaining decays of the second group, i.e.
$\Lambda \rightarrow n \gamma $,
$\Xi ^0 \rightarrow \Lambda \gamma $, and
$\Xi ^0 \rightarrow \Sigma ^0 \gamma $.

\subsection{Theoretical conflict}

Although any reasonable theoretical approach must have a built-in dominance of
two-quark transitions, 
such approaches may still differ in various ways.
The 
issue of {\em how} we take quark degrees of freedom into account 
lies at the origin of conflict
between these approaches. 
Namely, various models proposed may be classified into two groups
according to whether they satisfy or violate Hara's theorem.
In my opinion, 
models violating Hara's theorem should not be rejected immediately
in view of the fact that
1) we have already learned that the assumptions upon which Hara's theorem
is based, although seemingly correct for WRHD's, 
are not valid in Nature in general, and
2) experimental data seem to be better described by models violating
Hara's theorem (cf. Tables 1,4).
Among the approaches that satisfy Hara's theorem we should mention
the standard pole model of Gavela et al.\cite{GLOPR}, the chiral
perturbation theory framework \cite{Neufeld,Jenkins,BH99}, 
and the QCD sum rules
approach \cite{QCDK,QCDB}. Hara's theorem violating approaches
include simple quark-model calculations of Kamal and Riazuddin \cite{KR}
and the combined $VMD \times SU(6)_W$ approach of ref.\cite{Zen89,Zen91}
and its pole-model implementation \cite{Zen94}.

In order to analyze the issue of possible violation of Hara's theorem,
we should be able to compare experimental
asymmetries and branching ratios with their predictions in various models.
In principle, models might differ not only on the issue of whether Hara's
theorem is satisfied or violated (i.e. in the parity-violating amplitudes), 
but also in their description of the 
parity-conserving amplitudes.

\subsection{Parity-conserving amplitudes}
Clearly, if 
one wants to draw firm conclusions concerning  
parity-violating amplitudes on the basis of comparing theory 
with experiment, 
it is very important to use a reliable 
description of the parity-conserving amplitudes.
Fortunately, there are no real 
"conflicts" among various approaches to the latter.
Almost all papers agree here qualitatively, although they may differ
somewhat in their numerical predictions.
The most widely accepted approach is a hadron-level pole model, completely
analogous to that successfully used in the description of
nonleptonic hyperon decays (NLHD's). 
In this approach, quarks are used to find symmetry properties of two types of
hadronic blocks: 1) the amplitudes of photon emission by baryons, and
2) the amplitudes of weak transitions in baryons.
An alternative to that approach is to calculate the whole weak radiative
parity-conserving amplitude
at quark-level as one hadronic block, with no explicit intermediate
hadronic poles 
(using for example a bag model). 
Predictions of such an alternative approach do not differ qualitatively
from those of the pole model.
Since the pole model describes the data on NLHD's very well, and one does not
expect any physical complications (but rather simplification)
if the pion 
is replaced by a photon, it is reasonable to accept the pole model as a
reliable theoretical description of the parity-conserving WRHD amplitudes.

\subsection{Parity-violating amplitudes}
As in the case of parity-conserving amplitudes, the two-quark weak radiative
transition $su \rightarrow ud\gamma $ may be described either in terms
of several hadronic blocks, or as a single block.
Among many papers using the first approach one should mention first and foremost
the paper by Gavela, LeYaouanc, Oliver, Pene, and Raynal (GLOPR) \cite{GLOPR} 
in which a standard pole-model description of WRHD's is developed, and which 
provides a basis for any subsequent discussion on WRHD's.
This model satisfies Hara's theorem by construction.
The first group comprises also the chiral perturbation theory approach 
\cite{Neufeld,Jenkins,BH99}, and the Hara's-theorem-violating VMD-based
pole model of \cite{Zen94}.
The single-block approach was used in simple quark-model
calculations of Kamal and Riazuddin \cite{KR,VS}, in the bag model \cite{Lo},  
in the QCD sum rules approach \cite{QCDK,QCDB}, and in the combined
$SU(6)_W\times VMD$ approach of refs.\cite{Zen89,Zen91}.

\section{SPECIFIC MODELS AND THEIR PREDICTIONS}
\subsection{QCD sum rules}
QCD sum rules were applied to the description of WRHD's by Khatsimovsky
\cite{QCDK} and by Balitsky et al. \cite{QCDB}. Results of their calculations
are given in Table 2.  One can see that  
$\alpha (\Sigma \rightarrow p \gamma )$ is predicted to be positive, 
in complete
disagreement with the data (Table 1). The negative result of ref.\cite{QCDB}
was obtained only in a second attempt: the original calculation produced a 
positive sign (disguised as a negative one, due to a different
sign convention for asymmetry). 
Clearly, as agreed also by Khatsimovsky \cite{QCDK}, 
QCD sum rules do not have much 
predictive power.

\begin{table}
\caption{QCD sum rules: predictions  }
\begin{tabular}{lrr}
\hline
Decay & Asymmetry &Br. ratio$\cdot 10^3$\\
\hline
$\Sigma ^+ \rightarrow p \gamma$&$+1^{(1)}$&$0.8^{(1)}$\\
&$-0.85\pm0.15^{(2)}$&$0.5 ~{\rm to}~ 1.5^{(2)}$\\

$\Lambda \rightarrow n \gamma $&$+0.1^{(1)} $&$2.1-3.1^{(1)}$\\
$\Xi ^0 \rightarrow \Lambda \gamma $&$+0.9^{(1)}$&$1.1^{(1)}$\\
$\Xi ^-\rightarrow \Sigma ^- \gamma$ &$+0.4^{(1)}$&\\
\hline
\multicolumn{3}{l}
{$^{(1)}$ ref.\cite{QCDK}}\\
\multicolumn{3}{l}
{$^{(2)}$ 
ref.\cite{QCDB}, 
originally predicted positive}\\
\end{tabular}
\end{table}

\subsection{Chiral perturbation theory}
Attempts to describe WRHD's within chiral perturbation theory (ChPT) 
have not led to a
resolution of the problem. 
Ref.\cite{Jenkins} contains several free parameters
but the 
$\Sigma ^+ \rightarrow p \gamma$ asymmetry is still predicted to be small.
The analysis of Neufeld \cite{Neufeld} contains only
a small number of counterterms, and therefore has more predictive power.
Using as input the data on $\Xi ^0$ radiative decays available in 1992, 
ref.\cite{Neufeld} predicts then 
$|\alpha (\Sigma ^+ \rightarrow p \gamma)| < 0.2$,
$\alpha ( \Lambda \rightarrow n \gamma ) \approx -0.7 ~{\rm or}~-0.3$,
and $\alpha (\Xi ^-\rightarrow \Sigma ^- \gamma) \in (-0.4,+0.3)$.
The conclusion of Neufeld is that "the predictive power of ChPT
is limited by the occurrence of free parameters, which are not 
restricted by chiral (or other) symmetries alone".
In a recent paper \cite{BH99}, a new attempt to attack the issue within
a chiral approach has been made.  This approach is very similar to
the standard GLOPR paper because it is ultimately reduced to a pole model.
Therefore, it would be more appropriate to discuss it 
alongside ref.\cite{GLOPR}.
However, since the paper of ref.\cite{BH99} misses an important
contribution of intermediate $\Lambda (1405)$ \cite{Zen99BH}, 
its numerical predictions
for neutral hyperon decays have to be changed. It turns out \cite{Zen99BH}
that when this is done, one essentially recovers the predictions
of ref.\cite{GLOPR}.

\subsection{Standard pole model}
The standard approach of Gavela et al. 
\cite{GLOPR} was developed along the lines
of their earlier paper on NLHD's \cite{LOPR}.
Ref.\cite{LOPR} described parity-violating amplitudes of NLHD's as
composed of two terms: the current algebra commutator and a 
(vanishing in SU(3)) 
correction 
($\Delta P_{70}$) arising from $J^P=1/2^-$ intermediate states belonging 
to $(70,1^-)$ - the lowest-lying negative-parity multiplet of 
$SU(6)\times O(3)$, i.e. schematically:
\begin{equation}
\label{comm70}
A = [...,...] + \Delta P_{70}(m_s-m_{d})
\end{equation} 
with $\Delta P_{70}(0)=0$.

Alternatively, one might saturate the current algebra
commutator with this part of contribution from $(70,1^-)$
which does not vanish in $SU(3)$:
$[...,...]=P_{70}(0)$.
In other words, instead of the decomposition 
made on the right-hand side of Eq.(\ref{comm70}),
one might use a pole model with $SU(3)$ breaking
appropriately included: 
\begin{equation}
\label{polemodel}
A= P_{70}(m_s-m_d)=P_{70}(0)+\Delta P_{70}(m_s-m_{d})
\end{equation}
Diagrams relevant for this model are shown in Fig.1, where
$M$ stands for $\pi$ meson, and $B_{k*}$ - for all
allowed $J^P=1/2^-$ baryons from the $(70,1^-)$ multiplet.
If one wants to reproduce results of current algebra, one
has to consider {\em all} allowed negative parity baryons
from all SU(3) multiplets in $(70,1^-)$,
i.e. $\Lambda (1405) $ (a SU(3) singlet),
$N(1535)$, $\Lambda(1670)$, $\Sigma(1750)$ (low-lying SU(3) octet), etc.

\setlength{\unitlength}{0.55pt}
%--------DIAGRAMS --of --FIG.1
\begin{center}
\begin{picture}(260,500)

\put(0,300){
\begin{picture}(250,200)
\put(30,80){\line(-1,0){30}}
\put(70,80){\vector(-1,0){40}}
\put(125,80){\line(-1,0){55}}
\put(180,80){\vector(-1,0){55}}
\put(220,80){\line(-1,0){40}}
\put(250,80){\vector(-1,0){30}}
\multiput(70,80)(0,6){15}{\line(0,1){3}}
\put(180,80){\circle*{10}}
\put(180,50){\makebox(0,0)[b]{$H_{weak}$}}
\put(220,90){\makebox(0,0)[b]{$B_{i}$}}
\put(120,90){\makebox(0,0)[b]{$B_{k*}$}}
\put(30,90){\makebox(0,0)[b]{$B_{f}$}}
\put(90,150){\makebox(0,0)[b]{$M$}}
\put(135,0){\makebox(0,0)[b]{$(1)$}}
\end{picture}
}
%------------------------
\put(0,80){
\begin{picture}(250,200)
\put(30,80){\line(-1,0){30}}
\put(70,80){\vector(-1,0){40}}
\put(125,80){\line(-1,0){55}}
\put(180,80){\vector(-1,0){55}}
\put(220,80){\line(-1,0){40}}
\put(250,80){\vector(-1,0){30}}
\multiput(180,80)(0,6){15}{\line(0,1){3}}
\put(70,80){\circle*{10}}
\put(70,50){\makebox(0,0)[b]{$H_{weak}$}}
\put(220,90){\makebox(0,0)[b]{$B_{i}$}}
\put(120,90){\makebox(0,0)[b]{$B_{k*}$}}
\put(30,90){\makebox(0,0)[b]{$B_{f}$}}
\put(200,150){\makebox(0,0)[b]{$M$}}
\put(135,0){\makebox(0,0)[b]{$(2)$}}
\end{picture}
}

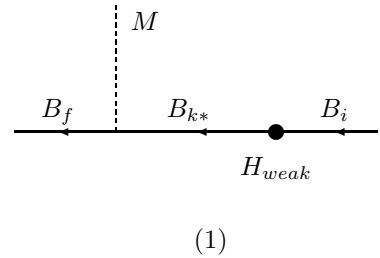
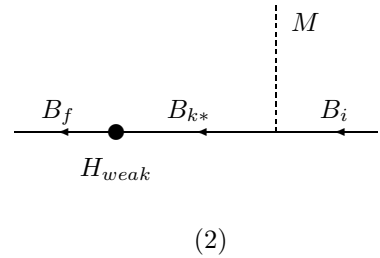
\put(130,15){\makebox(0,0)[b]{Fig.1. Baryon-pole diagrams}}
\end{picture}
\end{center}

For WRHD's, ref.\cite{GLOPR} switches to the pole model description.
This should give both an analogue of the commutator term for NLHD's
and the SU(3) breaking corrections.

%-------DIAGRAMS  -- of -- FIG.2-----------------------
\setlength{\unitlength}{0.45pt}
\begin{center}
\begin{picture}(432,350)
\put(1,20){
\begin{picture}(430,320)

%diagram (EM)
\put(230,160){
\begin{picture}(200,150)
\put(30,90){\line(1,0){25}}
\put(100,90){\vector(-1,0){45}}
\put(100,90){\line(1,0){45}}
\put(170,90){\vector(-1,0){25}}

\multiput(95,90)(-10,10){5}{\put(0,0) {\oval(10,10)[tr]}
                             \put(0,10){\oval(10,10)[bl]}}

\put(170,65){\vector(-1,0){70}}
\put(100,65){\line(-1,0){70}}
\put(170,40){\vector(-1,0){70}}
\put(100,40){\line(-1,0){70}}
\multiput(85,90)(5,0){6}{\line(1,0){3}}
\end{picture}}
%--------------------------
%diagram EM hadron
\put(0,160){
\begin{picture}(200,150)
\put(170,65){\vector(-1,0){35}}
\put(135,65){\line(-1,0){35}}
\put(100,65){\vector(-1,0){35}}
\put(65,65){\line(-1,0){35}}
\multiput(95,65)(-10,10){5}{\put(0,0) {\oval(10,10)[tr]}
                             \put(0,10){\oval(10,10)[bl]}}

\put(160,50){\makebox(0,0){k*}}
\put(40,50){\makebox(0,0){f}}

\end{picture}}
\put(220,225){\makebox(0,0){=}}
%--------------------------

%diagram (WEAK)
\put(230,0){
\begin{picture}(200,150)

\put(170,90){\vector(-1,0){70}}
\put(100,90){\line(-1,0){70}}

\put(170,65){\vector(-1,0){70}}
\put(100,65){\line(-1,0){70}}
\put(170,40){\vector(-1,0){70}}
\put(100,40){\line(-1,0){70}}
\multiput(100,40)(0,5){5}{\line(0,1){3}}
\put(105,45){$W$}
\end{picture}}
%--------------------------
\put(220,65){\makebox(0,0){=}}
%diagram (WEAK hadron)
\put(0,0){
\begin{picture}(200,150)

\put(160,50){\makebox(0,0){i}}
\put(40,50){\makebox(0,0){k*}}

\put(170,65){\vector(-1,0){35}}
\put(135,65){\line(-1,0){35}}
\put(100,65){\vector(-1,0){35}}
\put(65,65){\line(-1,0){35}}

\put(100,65){\circle*{10}}

\end{picture}}
%--------------------------------
\end{picture}}%-----of Fig.1
\end{picture}
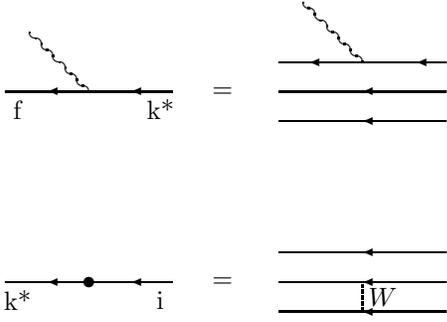
Fig.2 Hadron-level diagrams 
and their quark-level counterparts

\end{center}

The procedure applied in ref.\cite{GLOPR} is as follows:

1) Use quark model to evaluate symmetry properties of the two
(weak and electromagnetic) hadronic blocks in diagram (1) of Fig.1
with $M$ now replaced by $\gamma $ (as shown in Fig.2).

2) Determine the amplitudes for diagram (2) in Fig.1 from hermiticity,
CP- and gauge-invariance.

i) the weak amplitude is antisymmetric by $CP$ and hermiticity:
$a_{jk^*}=-a_{k^*j}$ ($j=i,f$).

ii) the obtained electromagnetic coupling is identified with
gauge-invariant hadron-level parity-conserving coupling
\begin{equation}
\label{f2}
f_{2,fk^*}\overline{u}_{1/2^+,f}\sigma ^{\mu \nu}
\gamma _5 q_{\nu } u_{1/2^-,k^*} A^{\mu }
\end{equation}
where $f_{2,jk^*}=f_{2,k^*j}$ by $CP$ and hermiticity.

By combining weak and electromagnetic transitions according to
Fig.1, one gets
\begin{eqnarray}
\label{GLOPRpv}
A&\propto &
\sum_{k^*}
\left\{
\frac{f_{2,fk^*}a_{k^*i}}{m_i-m_{k^*}}+
\frac{a_{fk^*}f_{2,k^*i}}{m_f-m_{k^*}}
\right\}\times \\
&&\times \overline{u}_fi\sigma ^{\mu \nu} \gamma _5 q_{\nu} u_i A_{\mu}
\end{eqnarray} 
For $i=f$ (which is almost the Hara's case) we use symmetry properties of
$a$ and $f_2$ to obtain:
\begin{equation}
f_{2,ik^*}a_{k^*i}=-a_{ik^*}f_{2,k^*i}
\end{equation}
which ensures cancellation of the first and second term in
Eq.(\ref{GLOPRpv}).
The sums (over $k^*$) of the first and second terms can be evaluated 
(in SU(6)) and
are given (in arbitrary normalization) in Table 3 (apart from the "$-$" sign
requested by symmetries of $a$ and $f_2$).
The prescription of the standard pole model is well defined and leads to definite
predictions for the signs of asymmetries (Table 4).
One obtains negative asymmetries for {\em all} four decays proceeding through
two-quark transitions. This $(-,-,-,-)$ pattern of asymmetries for 
$\Sigma^+$, $\Lambda $, and two $\Xi ^0$ decays is a characteristic
feature of the standard Hara's-theorem-satisfying model.
\begin{center}
\begin{table}
\caption{Weights of diagrams (1) and (2) of Fig.1 }
\vspace{1 mm}
\begin{tabular}{lrr}

\hline
&&\\[-3 mm]
Decay & Diagram (1) & Diagram (2)\\[1 mm]
\hline
&&\\[-2 mm]
\vspace{2 mm}
$\Sigma ^+ \rightarrow p \gamma$&$-\frac{1}{3\sqrt{2}}$&$-\frac{1}{3\sqrt{2}}$\\
\vspace{2 mm}
$\Lambda \rightarrow n \gamma $&$+\frac{1}{6\sqrt{3}}$&$+\frac{1}{2\sqrt{3}}$\\
\vspace{2 mm}
$\Xi ^0 \rightarrow \Lambda \gamma $&$0$&$-\frac{1}{3\sqrt{3}}$\\
\vspace{2 mm}
$\Xi ^0 \rightarrow \Sigma ^0 \gamma $&$\frac{1}{3}$&$0$\\

\hline
\end{tabular}
\end{table}
\end{center}

\subsection{"Naive" quark-level one-block calculation}
In 1983 Kamal and Riazuddin (KR) 
calculated $W$-exchange accompanied by photon radiation
in a simple quark framework \cite{KR}. The astonishing result of their
calculation was an explicit
violation of Hara's theorem (in the SU(3) limit). 
Although an agreement now exists that the calculation of ref.\cite{KR} is
completely correct from the technical point of view (\cite{Apr,Hpr,Z}, 
the disagreement still lingers as to the origin of the offending result and, 
consequently, how to treat it.

Azimov \cite{Azi97} proposed a way of proceeding 
if one identifies the result of KR calculations
with the $j^{(1)}_5$
 ($\gamma _{\mu}\gamma _5$-like) term in the full electromagnetic current
(if this term is present Hara's theorem may be violated - cf. section 3.1). 
He noticed that in principle 
the perturbative KR calculation may be supplemented
with a $\gamma _5$-dependent renormalization. 
Using the latter he showed that  
the $\gamma _{\mu}\gamma _5$-like 
term may be rotated away. 
In other words, one can "hide" the  
$\gamma _{\mu}\gamma _5$ term of the axial current
into the standard $\gamma _{\mu}$ 
piece of the vector current.  
This means that the concepts of left and right are redefined
in such a way that ultimately all the offending KR contribution constitutes a 
weak-interaction correction to the usual electromagnetic vector current.

The above idea may be applied to charged baryons only. 
In reality however, KR-like calculations may be performed for neutral baryons 
as well.  It turns out that the result is again non-zero.  This time, however,
this result (which conflicts with Hara-like considerations)  
cannot be rotated away 
since there is no $\gamma _{\mu}$ term in the vector current
of neutral baryons \cite{Zen98}.
One concludes \cite{Zen98} that 
the origin of KR result is completely unrelated to
the mechanism considered in ref.\cite{Azi97}.

In my opinion (shared by Holstein \cite{Hpr}), the result of KR is due
to the use of free quarks in states of definite momenta.  
This violates Hara's theorem 
because one of the theorem's assumptions is 
that we  deal  with a single object - 
a baryon in a state of definite momentum, and not with
a collection of free quarks.
This seems to mean that the KR result should be considered to be an artefact
of their model, and not a feature of reality \cite{Hpr}.

I think that the KR result is an artefact of their model if interpreted
literally: it
arises from free Dirac quarks propagating over infinite distances.
However, general features of the KR approach need not be 
incorrect.
The problem is that we still do not have a complete understanding
of how unobservable quarks combine to form such composite states as hadrons.
In the words of Donoghue et al. \cite{DGH}: 
"The quark model was developed in the first place to explain flavor
and spin properties of the observed hadrons and for this it does a good job.
The spatial aspect is less well tested."
It is precisely the question of position/momentum space description of
hadrons as quark composites that leads to the result of KR.

\subsection{Alternatives - bag model and VMD}

The quark model used by KR may be viewed as deficient.
Let us therefore accept for the time being that its result is an artefact.
Consequently, one has to replace the KR model with another, 
more "reasonable" approach.  This new approach should still exhibit 
spin-flavor symmetries that form the basis of all quark model successes,
but quarks should not be treated as free Dirac particles.
There are two possible ways of doing this: confining quarks to a bag
or using the idea of VMD combined with spin-flavor symmetries of hadrons.

Bag model calculations of Lo \cite{Lo}
show that the parity-violating
amplitude of the $\Sigma ^+ \rightarrow p \gamma $ is much larger than the
corresponding parity-conserving amplitude, again contradicting
expectations based on Hara's theorem.  Apparently, in bag model
calculations Hara's theorem still seems to be violated \cite{Z},
albeit the reasons are not clear and should be studied
more closely.
The bag model starts with the concept of free Dirac
quarks, and then confines them.
This proposes a resolution of the problem
by brute force of an additional assumption
and seems logically questionable to me: it assumes the answer.
I much prefer using the combined $VMD\times SU(6)_W$ approach, where questions
related to quark freedom or confinement are never asked, but
which "always works", although, admittedly, it is not completely clear why.
The approach does not use the concept of "free quarks" but yields quark model
results. Among its many successes one may mention here the 
successful prediction of baryon magnetic
moments by Schwinger \cite{Schwinger} (unlike in the 
constituent quark model, even the scale was predicted).  
It is also known that
a gauge-invariant formulation of the VMD approach is possible \cite{KLZ}.
An additional asset of the $VMD\times SU(6)_W$ 
approach as applied to WRHD's is that essentially all parameters are set by
NLHD's.
Thus, we are dealing with an easily falsifiable approach of great predictive
power.

The main idea of the VMD approach is as follows.
One starts with the standard SU(3) symmetric
model of parity-violating NLHD amplitudes
(Eq.\ref{comm70}) 
and uses spin-flavor $SU(6)_W$
symmetry 
to obtain weak strangeness-changing amplitudes for virtual transverse
vector meson (V)
emission from a baryon (B). This part is calculated following the ideas
of ref.\cite{DDH}. 
In this way, a transverse-vector-meson analogue of the commutator
term in Eq.(\ref{comm70}) is found \cite{Zen89}. 
In ref.\cite{DDH} it is identified with the $\gamma _{\mu} \gamma _5$
term in the general expression for the BBV amplitude.
The next step is to allow for standard VMD transition of vector meson 
into photon.
Thus, VMD suggests that transverse photon coupling to the 
electromagnetic axial weak current should proceed through the
$\gamma _{\mu} \gamma _5$ term.
Clearly, the conditions under which Hara's theorem was proved are not
satisfied now, and the approach chosen to avoid the use of free quarks 
(and the related problems with Hara's theorem) again 
exhibits its violation.
The parity-violating 
amplitudes of the VMD approach may be saturated with the contribution
from intermediate $J^P=1/2^-$ baryons, in a way completely analogous
to the case of NLHD's. 
The situation is similar to that occurring in the standard pole model
of Gavela et al. \cite{GLOPR}.
There is an important difference visualised in Fig.3, though.

%-------DIAGRAMS  -- of -- FIG.3-----------------------
\setlength{\unitlength}{0.45pt}
\begin{center}
\begin{picture}(432,250)
\put(1,40){
\begin{picture}(430,210)

%diagram (EM)
\put(230,0){
\begin{picture}(200,160)
\put(170,65){\vector(-1,0){35}}
\put(135,65){\line(-1,0){35}}
\put(100,65){\vector(-1,0){35}}
\put(65,65){\line(-1,0){35}}
%\multiput(95,65)(-10,10){5}{\put(0,0) {\oval(10,10)[tr]}
%                             \put(0,10){\oval(10,10)[bl]}}

\put(95,65){\line(-1,1){50}}
\put(95,50){\makebox(0,0){$\gamma _{\mu}\gamma _5$}}

\put(168,50){\makebox(0,0){k*}}
\put(26,50){\makebox(0,0){j}}
\put(40,120){\makebox(0,0){$V$}}
\end{picture}}
%--------------------------
%diagram EM hadron
\put(0,0){
\begin{picture}(200,160)
\put(170,65){\vector(-1,0){35}}
\put(135,65){\line(-1,0){35}}
\put(100,65){\vector(-1,0){35}}
\put(65,65){\line(-1,0){35}}
\multiput(95,65)(-10,10){5}{\put(0,0) {\oval(10,10)[tr]}
                             \put(0,10){\oval(10,10)[bl]}}

\put(168,50){\makebox(0,0){k*}}
\put(26,50){\makebox(0,0){j}}
\put(40,120){\makebox(0,0){$\gamma$}}

\put(92,50){\makebox(0,0){$\sigma _{\mu \nu}q^{\nu}\gamma _5$}}

\end{picture}}
\put(220,65){\makebox(0,0){$\rightarrow $}}
%--------------------------

%--------------------------------
\end{picture}}%-----of Fig.1
\end{picture}
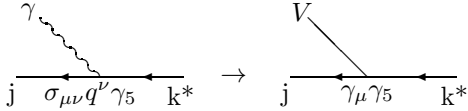
Fig.3 Photon emission in standard pole model and 
its vector meson counterpart

\end{center}

The difference is that $f_{2,jk^*}$, which accompanies the
$\sigma _{\mu \nu }q^{\nu}\gamma _5$ term (Eq.\ref{f2}), is symmetric under 
$j \leftrightarrow k^*$ interchange, while the $f_{1,jk^*}$ accompanying
the $\gamma _{\mu} \gamma _5$ vector-meson coupling is antisymmetric.
When one combines weak and electromagnetic transitions 
according to Fig.1 with $M=V$ and subsequently uses VMD, one obtains

\begin{eqnarray}
\label{VMDpv}
A&\propto &
\sum_{k^*}
\left\{
\frac{f_{1,fk^*}a_{k^*i}}{m_i-m_{k^*}}+
\frac{a_{fk^*}f_{1,k^*i}}{m_f-m_{k^*}}
\right\}\times \\
&&\times \overline{u}_fi\gamma ^{\mu} \gamma _5 u_i A_{\mu}
\end{eqnarray}
By using symmetry properties of $a$ and $f_1$
one finds that the term in brackets is now symmetric under
$i\leftrightarrow f$ interchange.
In other words, the two contributions in Table 3 add now rather than subtract.
An immediate consequence is that 
1) Hara's theorem is violated, and
2) asymmetries of the $\Lambda \rightarrow n \gamma $ and
$\Xi ^0 \rightarrow \Lambda \gamma $ are now positive.
The $(-,+,+,-) $ pattern obtained here 
for the $\Sigma ^+ \rightarrow p \gamma $, 
$\Lambda \rightarrow n \gamma $,
$\Xi ^0 \rightarrow \Lambda \gamma $,
and $\Xi ^0 \rightarrow \Sigma ^0 \gamma $ decays is a characteristic feature
of Hara's-theorem-violating approaches.
A comparison of asymmetry 
predictions of the VMD approach \cite{LZ}, the KR model \cite{VS}, and
the GLOPR standard pole model \cite{GLOPR} is given 
in Table 4.

\begin{table}
\caption{Model predictions  }
\begin{tabular}{lrrr}
\hline
&&&\\[-3 mm]
Decay &  VMD  & KR & GLOPR  \\[1 mm]
\hline
&&&\\[-3.5 mm]
$\Sigma ^+ \rightarrow p \gamma$  &$-0.95$ &$-0.56$
&$-0.80^{+0.32}_{-0.19}$\\[1 mm]

$\Lambda \rightarrow n \gamma $ &$+0.8$ &$-0.54$ &$-0.49$\\[1 mm]
$\Xi ^0 \rightarrow \Lambda \gamma $& $+0.8$& $+0.68$&$-0.78$\\[1 mm]
$\Xi ^0 \rightarrow \Sigma ^0 \gamma $&$-0.45$&$-0.94$&$-0.96$\\[1 mm]
\hline
\end{tabular}
\end{table}

From the comparison of model predictions with data (Table 1)
we see that at present the data favor approaches that violate Hara's theorem.
Asymmetry of the $\Xi ^0 \rightarrow \Lambda \gamma $ is crucial here.
It is large in all approaches, with its sign being negative (positive)
depending on whether Hara's theorem is satisfied (violated).
The fact that it is almost equal in absolute value in all approaches with great 
predictive power is not an accident.  
It can be traced directly to the sign of contribution
from diagram (2) and the vanishing of the contribution
from diagram (1) (Fig.1 and Table 3).  
The data point is three standard deviations away from the standard pole model.

Information from the comparison of asymmetries
is supplemented with that coming from branching ratios.
So far all data are best described by the VMD model \cite{Zen91,LZ}.

\section{SUMMARY}
The problem of WRHD's is already thirty years old. Data and some models
hint that Hara's theorem may be violated.
The KR result should certainly be treated as an artefact
if it were the only model which violates Hara's theorem.
However, other quark-inspired and elsewhere well-tested models
also violate the theorem, unless it is imposed 
by brute force of an additional assumption, foreign to quark approaches 
themselves.
Consequently, either our present models of how photons interact with 
quark composites are
incorrect or, as I believe, one should treat model hints seriously
and try to understand what they might mean.

The issue of Hara's theorem violation may be settled experimentally.
The crucial information should come from the sign of the
$\Xi ^0 \rightarrow \Lambda \gamma $ asymmetry. If this asymmetry is
large and negative, Hara's theorem is satisfied and one has 
to conclude that various hints
were misleading.
If, on the other hand,
this asymmetry is positive, one has to conclude that violation of Hara's
theorem is a feature of Nature.
This would mean that at least one of the assumptions of Hara's theorem
is violated.

The asumptions of CP-invariance and current conservation are
satisfied explicitly in the KR paper.
We have pointed out that in the KR paper 
violation of Hara's theorem results from
the fact that in these calculations baryons consist of free quarks in plane-wave
states of definite momenta. 
From the point of view of position space ,such states
contain terms with far-away quarks. It is from such
configurations that violation of Hara's theorem originates. 
This picture hints at
the assumption of locality as the one that is violated.
Hadron-level prescription (such as that 
of VMD) in which hadron is described by a local field may be also analysed 
from the point of view of position space.
The net result is that CP-invariant 
interaction of photon with a conserved baryonic axial current does lead
to the violation of Hara's theorem if
the current exhibits a kind of nonlocality \cite{ZenDm}.
Thus, although the detailed origin for the violation of Hara's theorem
is different in these two approaches, they both hint at nonlocality
as the potential culprit.
Since we know that nonlocality is a general feature
of composite quantum states,
the above conclusion is not in conflict with the general properties 
of the quantum world. 
However,
it is certainly weird as it does challenge the 
generally accepted simple pictures of hadrons and 
photon-hadron interactions.

Since, apart from the arguments and hints presented
in this talk, one can also invoke arguments of a much deeper, though usually 
disregarded kind,
I find it quite believable that Hara's theorem may be violated in Nature.

\end{document}